\documentstyle[floats,prl,aps,epsf,epsfig]{revtex}
\draft % this command allows pacs numbers to be printed
\begin{document}

  %%%%%%%%%%%%%%%%%%%%%%%%%%%%%%%%%%%%%%%%%%%%%%%%%%%%%%%%%%%%%%%%%%%%%%
  %
  %\renewcommand{\topfraction}{0.8}
   
   %  This is the first line to be uncommented for 2 column format
   \twocolumn[\hsize\textwidth\columnwidth\hsize\csname
   @twocolumnfalse\endcsname
   %
   %%%%%%%%%%%%%%%%%%%%%%%%%%%%%%%%%%%%%%%%%%%%%%%%%%%%%%%%%%%%%%%%%%%%%%

     %\title{Zero-point
     \title{
     Approximate black holes
     from a variable cosmological constant}
     \author{Maurice H.~P.~M. van Putten}
     \address{Massachusetts Institute of Technology,
	      Cambridge, MA 02139}
	       
	       \maketitle
		
		\begin{abstract}
     \mbox{}\\
     The small or zero cosmological constant, $\Lambda$,
     probably results from a macroscopic
     cancellation mechanism of the zero-point
     energies. 
     However, nearby horizon surfaces any macroscopic 
     mechanism is expected to result in
     imperfect cancellations. 
     A phenomenological description
     is given for the residual
     variable cosmological 
     constant. 
     In the static, spherically symmetric case it produces
     approximate black holes. 
     The model describes
     the case of exponential decay by
     $\Box\ln\Lambda=-3a$,
     were
     $a$ is a positive constant.
     \end{abstract}
 
  
  \vskip2pc]
   
   The modern saga of the cosmological
   constant problem concerns the
   divergence of the zero-point
   energies \cite{wein:a}.
   A particularly compelling mechanism
   for cancellation of the resulting
   cosmological constant has been proposed by Coleman
   in the Euclidean path integral
   approach \cite{co:a}.
   Coleman's analysis predicts a zero expectation value of
   the cosmological constant by
   integration over thin 
   but extended wormholes. 
   However, a horizon surface is 
   expected to perturb any such macroscopic cancellation mechanism,
   by suppression of those wormholes which would tend to
   penetrate the surface.
   This perturbation suggests the possibility for a finite
   cosmological constant to reappear in the vicinity of 
   horizon surfaces.
 In this $Letter$, a simple
 phenomenological approach
 is described to illustrate
 this scenario,
 using a self-consistent 
 classical description
 of a variable cosmological constant.
  
  A space-time with 
  finite cosmological constant,
  $\Lambda$, filled with a fluid
  with stress-energy tensor,
  $T_{\alpha\beta}$,
  gives rise to the equations of motion
  \begin{eqnarray}
  \begin{array}{rl}
  G_{\alpha\beta}=&-\Lambda
  g_{\alpha\beta}
  +S_{\alpha\beta}\equiv T_{\alpha\beta}.
  \end{array}
  \label{EQN_EE}
  \end{eqnarray}  
 We shall look for 
 time-independent, spherically
 symmetric solutions in
 the Schwarzschild line-element
 \begin{eqnarray}
 \mbox{d}s^2=e^\lambda\mbox{d}r^2
	     +r^2(\mbox{d}\theta^2+
			    \sin^2\theta\mbox{d}
					     \phi^2)-e^\nu\mbox{d}t^2.
					     \end{eqnarray}

A spherical region with finite 
$\Lambda$
must be suitably closed by a shell
with positive hoop stress to sustain 
the negative (radial) pressure inside.
The shell also resides
in a deep gravitational potential well, and therefore
is expected to accumulate radiation
and particles. 
If the particles interact infrequently
(relative to the orbital time-scale),
the stress tensor will develop
an anisotropic pressure 
dominated by
random orbital motion.
In the limit of zero radial
pressure of the fluid,
its stress-energy tensor 
becomes 
\begin{eqnarray}
S_{\alpha\beta}=
(E+P)u_\alpha u_\beta+
Pk_{\alpha\beta},
\end{eqnarray}
where $u^\beta=(0,0,0,e^{-\nu/2})$ is the
the average velocity 
four-vector of the particles, 
\begin{eqnarray}
k_{\alpha\beta}=g_{\alpha\beta}-
(\partial^\gamma\Lambda\partial_\gamma\Lambda)^{-1}
\partial_\alpha\Lambda\partial_\beta\Lambda
\end{eqnarray}
is a projection tensor,
$E$ is the energy density
of the fluid and $P$ is its
two-dimensional pressure. Hence, we have
\begin{eqnarray}
T_\alpha^\beta=
-\mbox{diag}(\Lambda,M,M,\Lambda+E),
\end{eqnarray}
were we have set $M=\Lambda-P$.
It may be seen that the quantities
$\Lambda, E$ and $P$
are scalars.
The equations of 
energy-momentum 
conservation, 
\begin{eqnarray}
\nabla_\alpha
T^\alpha_\beta=
\partial_\alpha T^\alpha_\beta
+\Gamma^\alpha_{\alpha\gamma}
T^{\gamma}_{\beta}-\Gamma_{\alpha\beta}^\gamma
T_\gamma^\alpha=0,
\end{eqnarray}
reduce to 
\begin{eqnarray}
-\Lambda^\prime+\Gamma^\alpha_{\alpha r}
T^r_r-\Sigma_\alpha \Gamma^\alpha_{\alpha r}
T^\alpha_\alpha=0.
\end{eqnarray}
With the tabulated expressions
for the connections
(e.g.: \cite{steph:a}), this
obtains
\begin{eqnarray}
\Lambda^\prime=\frac{2}{r}(M-\Lambda)+\frac{\nu^\prime}{2}E.
\label{EQN_L}
\end{eqnarray}
At this point, 
recall the
the 
weak energy condition on
the stress-energy tensor
\begin{eqnarray}
\mbox{inf}~
T_{\alpha\beta}
\xi^\alpha\xi^\beta=
\Lambda+E\ge0,
\label{EQN_P}
\end{eqnarray}
were the infimum is taken
over all time-like
vectors $\xi^\beta$.
Clearly, (\ref{EQN_P})
allows for the fluid to
consist not only of
particles
($E_p>0$), but also
of negative mass holes
($E_h<0$) in regions with
$\Lambda>0$. 
Thus, coexistence of 
particles and holes
makes it possible to have
positive
hoop stress,
even with $E=0$. 

     We will look for
     Schwarzschild solutions which join to 
     space-time
     with a variable cosmological
     constant in the neighborhood
     of the origin, that is
     \begin{eqnarray}
     e^\lambda\sim&
     \left(1-\frac{\Lambda_0r^2}{3}
     \right)^{-1},~
     e^\nu\sim1-\frac{\Lambda_0r^2}{3},
     ~E,P\sim0,
     \end{eqnarray}
were $\Lambda_0$ is the value of the
cosmological constant at the origin.
The relevant Einstein equations
are those for $G^r_r$ and
$G^t_t$, 
\begin{eqnarray}
\begin{array}{rl}
e^{-\lambda}(\frac{\nu^\prime}{r}
+\frac{1}{r^2})-\frac{1}{r^2}
=&
-\Lambda,\\
-e^{-\lambda}
	  (\frac{\lambda^\prime}{r}-\frac{1}{r^2})
	  -\frac{1}{r^2}
=&
-(\Lambda+E).
\end{array}
\label{EQN_GR}
\end{eqnarray}
 These may be written
 as
\begin{eqnarray}
\begin{array}{rl}
r\nu^\prime=&-1+e^\lambda(1-r^2\Lambda),\\
r\lambda^\prime=&
1-e^\lambda
(1-r^2[\Lambda+E]).
\end{array}
\label{EQN_GR2}
\end{eqnarray}
Whenever
$\Lambda=E=P=0$
beyond 
some $r=r_0$,
(\ref{EQN_GR2}) recovers
the Schwarzschild solution,
though perhaps
after a rescaling of the time
coordinate (in case
of $\lambda+\nu\ne0$)
to maintain the familiar
relationship $-g_{tt}g_{rr}=1$.
Clearly, the simplest case
follows with
$E=0$, which is allowed in
our particle-hole model if
any gravitational separation
of particles and holes can
be neglected on the scale
of variation in $\Lambda$.
  
The simplest case, therefore,
is given by
a constitutive relation
only for the 
anisotropy, $P$, 
in the pressure.
The regularity
condition of $P$ at the origin,
$P=O(r^2)$ for small
$r$, translates covariantly
to \begin{eqnarray}
P=\frac{\partial^\gamma\Lambda
\partial_\gamma\Lambda}{F(\lambda)},
\label{EQN_PP}
\end{eqnarray}
were $F(\Lambda)$ is 
a continuous function of 
$\Lambda$. 
For example,
$F(\lambda)=2a\Lambda^\alpha$,
so that
\begin{eqnarray}
P=\frac{e^{-\lambda}}{2a\Lambda^\alpha}
(\Lambda^\prime)^2
\end{eqnarray}
giving a reduction of
(\ref{EQN_L}) to
\begin{eqnarray}
\Lambda^\prime=-a
re^\lambda\Lambda^\alpha.
\label{EQN_PL}
\end{eqnarray}
Equation (\ref{EQN_PL})
shows a 
positioning of a
transition layer in
$\Lambda$ at
large $e^\lambda$,
$i.e.$, at a surface of
large red shift.
In the case of $\alpha=1$, the covariant
generalization of 
(\ref{EQN_PL}) is
\begin{eqnarray}
\Box \ln\Lambda = -3a.
\label{EQN_LA}
\end{eqnarray}
A solution is illustrated in 
Figure 1. With
$0<\alpha<1$,
the distinguishing
feature arises of 
$\Lambda$ vanishing at finite
$r$, as illustrated in
Figure 2. 
The covariant generalization of
(\ref{EQN_PP}) for $\alpha\ne1$
is given by
\begin{eqnarray}
\Box \Lambda^{(1-\alpha)} = -3a
	     (1-\alpha); 
\label{EQN_LB}
\end{eqnarray}
for $\alpha>1$, $\Lambda(r)$ 
decays algebraically in the
far field limit.

\begin{center}
\epsfig{file=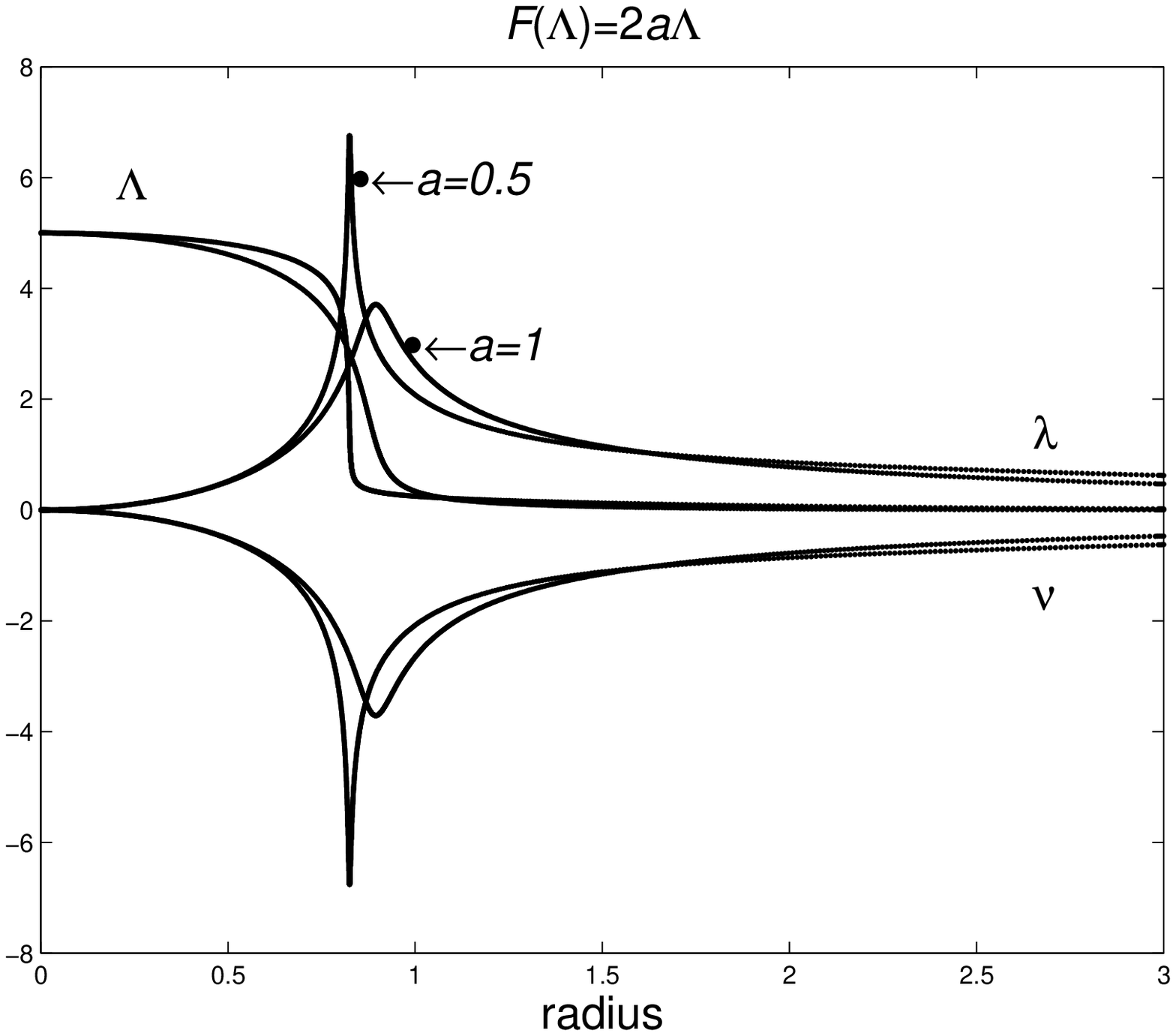,
width=8.5cm,height=6cm}
\end{center}
{\bf Figure 1.} 
{\small Distributions of the
metric quantities $\lambda$, $\nu$ and
$\Lambda$ as a function of 
radius in the case of 
$F=2a\Lambda$ ($\alpha=1$) for
$a=0.5,1$ and
$\Lambda=5$ at the
origin.
A high red shift accompanies
the transition layer in
$\Lambda$.
Outside of this layer, the
cosmological constant is exponentially
small,
and space-time is 
essentially
described
by the exterior Schwarzschild 
solution.
Inside this layer, the
cosmological constant is
effectively constant. The origin
is singulaty free.}
\mbox{}\\
\begin{center}
\epsfig{file=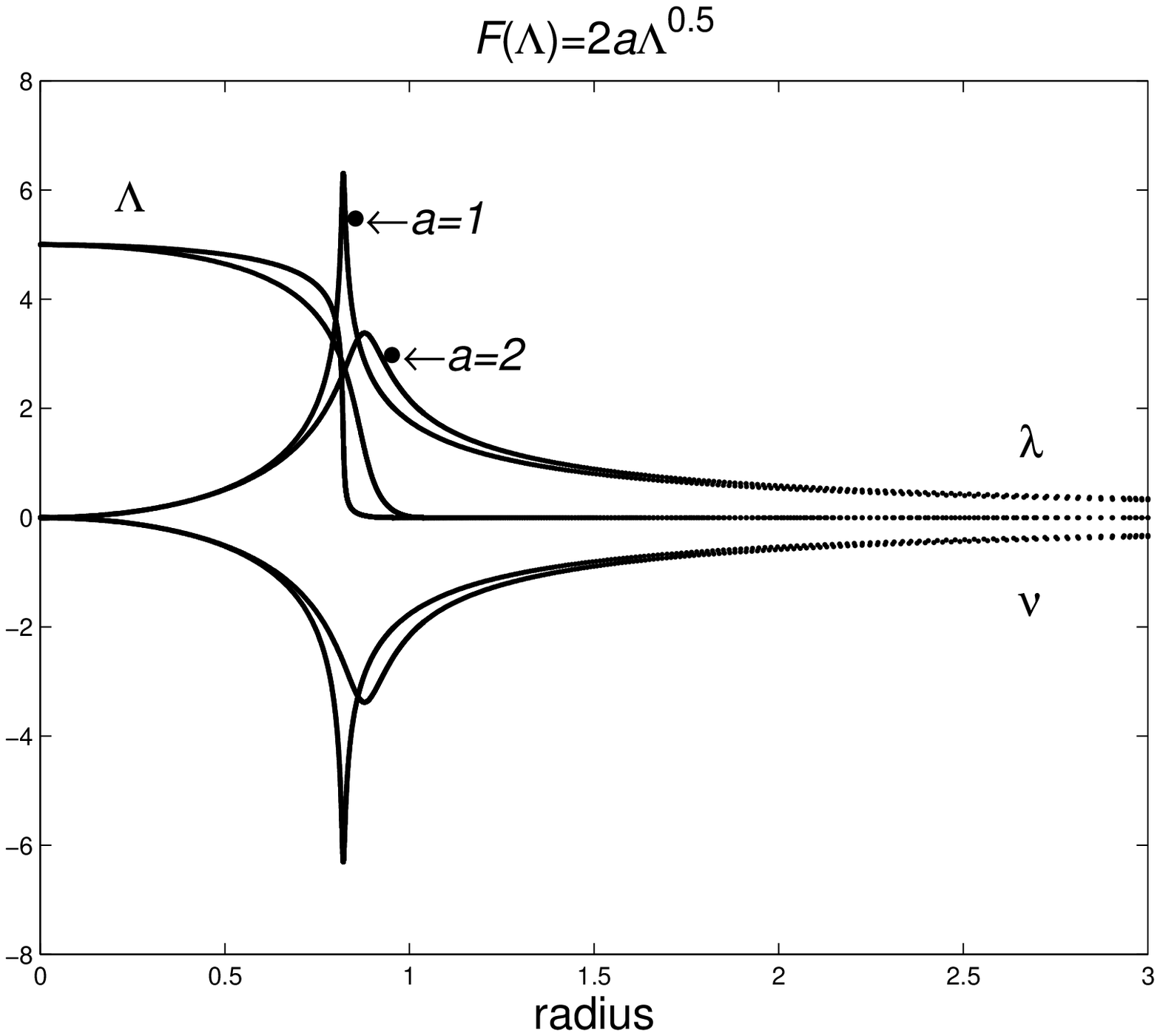,
width=8.5cm,height=6cm}
\end{center}
{\bf Figure 2.} 
{\small Distributions of the
metric quantities $\lambda$, $\nu$ and
$\Lambda$ as a function of 
radius in the case of 
$F=2a\Lambda^{0.5}$ ($\alpha=0.5$)
for $a=1,2$ and $\Lambda=5$ at the
origin.
The transition layer is
sharper compared with
the previous case
$F=2a\Lambda$ 
with the same $a=1$ as displayed
in Figure 1. Furthermore, 
a bifurcation surface appears
beyond which
$\Lambda$ vanishes
identically, 
slightly outside 
the surface of high
red shift. The origin is
singularity free.}
\mbox{}\\

It is an open question whether 
(\ref{EQN_LB}) preserves
non-negativity of 
$\Lambda^{(1-\alpha)}$ 
($\Lambda_0>0$) 
in dynamical space-times.
It may take numerical
simulations to study
the behavior of the
surface of bifurcation
(at which $\Lambda$ becomes 
zero) 
in this setting.
For this reason,
a preliminary preference
may be given to 
(\ref{EQN_LA}) which has
$\alpha=1$.
It should be metioned
that any terms on the
right hand-side of 
(\ref{EQN_LA}-\ref{EQN_LB})
which vanish for $E=0$ have
been suppressed.

The existence of 
solutions 
can be seen from boundedness of
$\lambda$. (\ref{EQN_GR2}) shows
that $\lambda$ can blow up at a
finite radius only
when $r^2\Lambda>1$.
By (\ref{EQN_GR2}),
$r\lambda^\prime\ge-(e^\lambda-1)$,
so that $\lambda\ge0$. Then
$\lambda^\prime\le re^\lambda\Lambda$
by the same equation.
By (\ref{EQN_PL}),
$re^\lambda\Lambda$ can be expressed
in terms of $\Lambda^\prime$, so that
\begin{eqnarray}
\lambda^\prime\le-\frac{1}{a}\frac{\Lambda^\prime}
{\Lambda^{\alpha-1}}.
\end{eqnarray}
It follows that 
\begin{eqnarray}
\lambda(r)\le\lambda(r_0)+\frac{1}{a}
\int_{\Lambda(r)}^{\Lambda(r_0)}
\frac{d\Lambda}{\Lambda^{\alpha-1}}.
\end{eqnarray}
Clearly, if $0<\alpha<2$, we may take $r_0=0$, so that
\begin{eqnarray}
0\le\lambda\le\frac{1}{a(2-\alpha)}\Lambda_0^{2-\alpha}<\infty.
\end{eqnarray}
For $\alpha\ge2$, choose $r_0$ such that
$\Lambda(r_0)r_0^2\ge1$. For those $r\ge r_0$, where
$\Lambda(r)r^2\ge1$, we have
\begin{eqnarray}
\begin{array}{rll}
\lambda(r)&\le&\lambda(r_0)+\frac{1}{a}
               \int_{1/r^2}^{\Lambda(r_0)}
               \Lambda^{1-\alpha}d\Lambda\\
          &\le&\lambda(r_0)+\frac{1}{a}\left\{
\begin{array}{lr}
\ln\Lambda(r_0)r^2&\small{(\alpha=2)},\\
r^{2(\alpha-2)}   &\small{(\alpha>2)}.
\end{array}
\right.
\end{array}
\end{eqnarray}
Again, no blow-up at finite
radius occurs.

The fully covariant
stress-energy tensor is
\begin{eqnarray}
S_{\alpha\beta}
 =Eu_\alpha u_\beta
   +Ph_{\alpha\beta}
  -\frac{\partial_\alpha\Lambda
   \partial_\beta\Lambda}{F(\Lambda)}
\label{EQN_S}
\end{eqnarray}
 with
  $h_{\alpha\beta}=
  g_{\alpha\beta}+
 u_\alpha u_\beta.$
The energy $E$ has been
included 
for consistency in 
view of the normalization
$u^2=-1$, so that
the four equations of 
energy-momentum conservation,
\begin{eqnarray}
\nabla_\alpha T^\alpha_\beta=0
\end{eqnarray}
can be satisfied 
with the three degrees of
freedom in $u^\beta$ plus the
one degree of freedom in
$E$. As we have seen, $E$
will depart from zero only
in genuinely dynamical space-times.
The evolution equation for
$E$ and $u^\beta$ satisfies
\begin{eqnarray}
\nabla_\alpha(Eu^\alpha u_\beta)
+\nabla_\alpha S^\alpha_\beta-\partial_\beta\Lambda=0;
\end{eqnarray}
in particular,
\begin{eqnarray}
\nabla_\alpha(Eu^\alpha)=
u^\beta\partial_\beta\Lambda+u^\beta
\nabla_\alpha S^\alpha_\beta
\label{EQN_E}
\end{eqnarray}
expresses the
evolution of $E$.

The present solutions 
feature surfaces of 
arbitrarily high 
red shift, outside of
which space-time is 
essentially that of
general relativity. 
These solutions
are related to the earlier
solution of van Putten
\cite{mvp:a}, which were
proposed as
as approximate black holes
for numerical regularization.
Spherically symmetric
static solutions with
surfaces of arbitrarily 
high red shift have recently
also been described 
by `t Hooft 
\cite{thooft:a,thooft:b},
in an entirely different
context. These 
solutions
are both hampered by naked singularities
at the origin
(though of different kinds). 
Numerical simulations
by Coptuik $et$ $al.$
\cite{chop:a}
also show the
solution of \cite{mvp:a}
to be unstable.
This naturally raises the
same question of 
stability of the
present solutions, something
which falls outside the scope
of the present work.
It should be emphasized
that the present solutions 
are an improvement 
(over forementioned related
solutions) in 
that these are free of any
singularities. 

In regards to the application
to numerical relativity, note that
the stress-energy tensor remains
differentiable
everywhere, though with
a discontinuity in its first 
derivative if $0<\alpha<1$. 
This still provides a proper
source term in
the covariant,
strictly hyperbolic formulation 
in terms of nonlinear
wave equations by van Putten 
\& Eardley \cite{mvp:b},
given by
\begin{eqnarray}
\hat{\Box}\omega_{\alpha ab}
-R^\gamma_\alpha\omega_{\gamma ab}
-[\omega^\gamma,\nabla_\alpha\omega_\gamma]_{ab}
=16\pi\tau_{\alpha ab},
\label{EQN_OM}
\end{eqnarray}
were $\omega_{\alpha ab}$ are the
tetrad connections of the tetrad
elements $(e_a)_\alpha$ (here, Latin
indices refer to the tetrad indices
and Greek indices refer to the coordinates).
The source term 
$\tau_{\alpha ab}=\tau_{\alpha\beta\gamma}
(e_a)^\beta(e_b)^\gamma$,
is given by
\begin{eqnarray}
\tau_{\alpha \beta\gamma}=
\nabla_{[\beta}T_{\gamma]\alpha}-
\frac{1}{2}
g_{\alpha[\gamma}\nabla_{\beta]}T,
\end{eqnarray}
were $T=T^\gamma_\gamma$ is the trace of the
stress-energy tensor. The
equations in the exterior vacuum, 
therefore, are
\begin{eqnarray}
\hat{\Box}\omega_{\alpha ab}
-[\omega^\gamma ,\nabla_\alpha\omega_\gamma ]_{ab}
=0.
\end{eqnarray}
These have been given
a second-order numerical implementation
in one dimension \cite{mvp:c}.
In the case of
$0<\alpha<1$,
$\tau_{\alpha ab}$ in (\ref{EQN_OM})
is at most discontinuous, due to second
derivatives of $\Lambda$;
for $\alpha\ge1$, there are 
no discontinuities. 
It is
expected, therefore,
that 
for all $\alpha>0$
the stress-energy tensor
(\ref{EQN_S}),
plus $-\Lambda 
g_{\alpha\beta}$
and constitutive
relation (\ref{EQN_PP}),
can indeed be integrated into this
approach 
of numerical
relativity.

\mbox{}\\
\centerline{\bf ACKNOWLEDGEMENTS}
\mbox{}\\
The author gratefully acknowledges stimulating
discussions with Ira Wasserman in the very
early stages of this work, Thomas
Witelski, Hung Cheng, 
Daniel Freedman and 
Alan Guth; and S. Caveny for a careful
reading. This work 
has received partial support
from 
the MIT Sloan/Cabot
Fund.
%\begin{references}

%\end{references}
   
   \end{document}